\documentclass{emulateapj}

\shorttitle{The Eclipsing white dwarf binary CSS 41177}
\shortauthors{S. G. Parsons et al.}

\begin{document}

\title{A deeply eclipsing detached double helium white dwarf binary}

\author{S. G. Parsons, T. R. Marsh and B. T. G{\"a}nsicke}
\affil{Department of Physics, University of Warwick, Coventry, CV4 7AL}

\author{A. J. Drake}
\affil{California Institute of Technology, 1200 E. California Blvd, CA 91225,
  USA}

\and

\author{D. Koester}
\affil{Institut f{\"u}r Theoretische Physik und Astrophysik, Universit{\"a}t Kiel, Germany}

\begin{abstract}
Using Liverpool Telescope+RISE photometry we identify the 2.78 hour period binary star CSS 41177 as a detached eclipsing double white dwarf binary with a 21,100K primary star and a 10,500K secondary star. This makes CSS 41177 only the second known eclipsing double white dwarf binary after NLTT 11748. The 2 minute long primary eclipse is $40\%$ deep and the secondary eclipse $10\%$ deep. From Gemini+GMOS spectroscopy we measure the radial velocities of both components of the binary from the H$\alpha$ absorption line cores. These measurements, combined with the light curve information, yield white dwarf masses of M$_1 = 0.283\pm0.064$M$_{\sun}$ and M$_2 = 0.274\pm0.034$M$_{\sun}$, making them both helium core white dwarfs. As an eclipsing, double-lined spectroscopic binary CSS 41177 is ideally suited to measuring precise, model-independent masses and radii. The two white dwarfs will merge in roughly 1.1 Gyr to form a single sdB star.
\end{abstract}


\section{Introduction}

There are of order 100--300 million close pairs of white dwarfs in the Galaxy. They are thought to be the dominant source of background gravitational waves detectable by {\it LISA} \citep{hils90}. The loss of orbital angular momentum via gravitational radiation in these systems will eventually bring the two white dwarfs into contact with one another. Those that achieve this within a Hubble time are possible progenitors of AM CVn binaries, R CrB stars and Type Ia supernovae \citep{iben84}. The coalescence of two low-mass, helium core white dwarfs is also a possible formation channel for the creation of single sdB stars \citep{webbink84}; indeed it may be one of the more important sdB formation channels \citep{han03}. There are now 43 detached double white dwarf binaries known (\citealt{kilic11}; \citealt{marsh11} and references therein) but few of these yield the precise parameters needed to determine their eventual fate.

The subject of this paper, CSS 41177 (SDSS J100559.10+224932.2), was listed as a white dwarf plus main sequence binary by \citet{drake10}. They used a marginal 2MASS detection to infer the presence of a late-type companion though they note that a small faint object could also produce the observed transit signal. We observed CSS 41177 as part of a monitoring campaign for period variations in white dwarf plus main sequence binaries. With a better sampled eclipse and detection of the secondary eclipse we realised that it was in fact a detached eclipsing double white dwarf binary, making it only the second known after NLTT 11748 \citep{steinfadt10}, but in this case the secondary star contributes 22\% of the overall flux (compared to 3\% in NLTT 11748). Follow up spectroscopy revealed that CSS 41177 was also a double-lined spectroscopic binary.

Here we present Liverpool Telescope (LT)+RISE photometry and Gemini+GMOS spectroscopy of CSS 41177, and use these observations, and mass-radius relations, to determine the system parameters. We also discuss the evolution and eventual fate of the binary.

\section{Observations and their reduction}

\subsection{LT+RISE photometry}
Three primary eclipses and two secondary eclipses of CSS 41177 were observed using the LT and RISE \citep{steele04} between February and April 2011. RISE is a high-speed frame transfer CCD camera with a single wideband V+R filter \cite{steele08}. All observations were made in $2 \times 2$ binning mode, giving a scale of 1.1 arcsec per pixel, and with exposure times of 12-13 seconds.

The raw data are automatically run through a pipeline that debiases, removes a scaled dark frame and flat-fields the data. The source flux was determined with aperture photometry using a variable aperture, whereby the radius of the aperture is scaled according to the FWHM, using the ULTRACAM pipeline \citep{dhillon07}. Variations in observing conditions were accounted for by determining the flux relative to the nearby star G 54-11 (V=14.6).

\begin{figure*}
\centering
\epsscale{.99}
\plotone{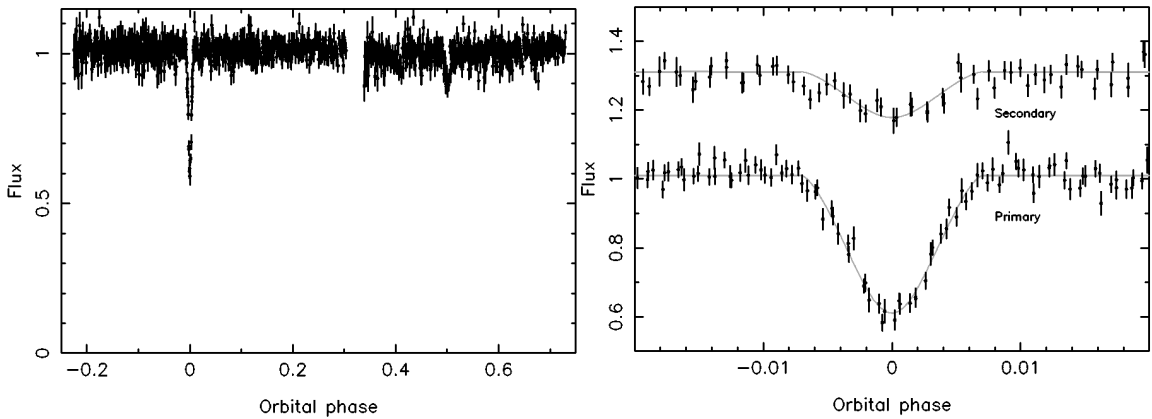}
\caption{\emph{Left:} LT+RISE photometry of CSS 41177 folded on its 2.78 hour orbital period and binned by a factor of two. \emph{Right:} The unbinned primary and secondary eclipse data. The secondary eclipse is shifted back by half a cycle and vertically by 0.3.\label{RISEfit}}
\end{figure*}

\subsection{Gemini+GMOS spectroscopy}

CSS 41177 was observed using GMOS on the 8m Gemini Observatory North telescope \citep{hook04} on 15 March 2011. We used the R831 grating with a 0.5 arcsec slit and $4\times2$ binning (spatial by spectral) to give a resolution of $R\sim4400$ centered at H$\alpha$. We used an exposure time of 10 minutes, recording 17 spectra covering an entire orbit.

The average bias and flat-field correction was carried out using the {\sc figaro} package from {\sc starlink} and nightly bias and tungsten frames. We optimally extracted each spectrum \citep{marsh89} and used a CuAr arc lamp to establish an accurate wavelength scale. The arc spectrum was fitted with a fifth-order polynomial to 15 lines which gave an rms of 0.05 pixel. The individual spectra were normalised to the continuum level using a polynomial fit to the continuum regions and the spectra were placed onto a heliocentric wavelength scale. 

\section{Results}

\subsection{Light curve model fitting}  \label{lcurve}

The left hand panel of Figure~\ref{RISEfit} shows the LT+RISE photometry of CSS 41177 folded on its 2.78 hour orbital period and binned by a factor of two. Since both eclipses are detected, the light curve information alone constrains the system inclination and the radii scaled by the binary separation.

To measure the system parameters we fitted the light curve using a code written to produce models for the general case of binaries containing a white dwarf (see \citealt{copperwheat10} for details). The program subdivides each star into small elements with a geometry fixed by its radius as measured along the direction of centres towards the other star. From an initial set of parameters defined by the user, the code produces model light curves which are initially fitted to the RISE data using Markov chain Monte Carlo (MCMC) minimisation to produce a set of covariances. We then ran another MCMC minimisation using these covariances to define the parameter jumps. To produce the final parameters and their errors we followed the procedures described in \citet{collier07}. The parameters needed to define the model were: the mass ratio, $q = M_\mathrm{2}/M_\mathrm{1}$, the inclination, $i$, the radii scaled by the binary separation, $R_\mathrm{1}/a$ and $R_\mathrm{2}/a$, the surface brightness ratio, $S_1/S_2$, quadratic limb darkening coefficients for the both white dwarfs, the time of mid eclipse, $T_{0}$ and the period, $P$.

We corrected all the times to Barycentric Dynamical Time (TDB). We initially fitted each light curve individually in order to measure mid-eclipse times for the three primary eclipses recorded.  The second eclipse recorded was set to cycle number zero. Our measured eclipse times are shown in Table~\ref{ecl_times}. The zero-point of the ephemeris and the period were allowed to vary in our final fit. We kept the mass ratio fixed as 1.0 (the light curves of this well-detached system are independent of the mass ratio). We determined the quadratic limb darkening coefficients from a pair of white dwarf atmosphere models with temperatures of 21,100K and 10,500K and $\log{g} = 7.3$ (see Section~\ref{temps}) folded through the RISE filter profile \citep{gansicke95}. We determined coefficients of a = 0.105 and b = 0.228 for the primary and a = 0.176 and b = 0.288 for the secondary for $I(\mu)/I(1) = 1-a(1-\mu)-b(1-\mu)^2$, where $\mu$ is the cosine of the angle between the line of sight and the surface normal; we kept these values fixed. The inclination, scaled radii and surface brightness ratio were allowed to vary.

\begin{table}
\begin{center}
\caption{Mid-eclipse times.\label{ecl_times}}
\begin{tabular}{@{}lcc@{}}
\tableline\tableline
Cycle  & Eclipse time & Uncert    \\
number & MJD(BTDB)    & MJD(BTDB) \\
\tableline
-171 & 55599.087788   & 0.000013 \\
0    & 55618.926463   & 0.000016 \\
302  & 55653.963116   & 0.000020 \\
\tableline
\end{tabular}
\end{center}
\end{table}

The right hand panel of Figure~\ref{RISEfit} shows the best fit to the RISE photometry around the eclipses. The two eclipses constrain the system inclination to $89.2^\circ\pm0.3^\circ$. The constraints on the two scaled radii are shown in the left hand panel of Figure~\ref{constrain}. We also update the ephemeris to
\[\mathrm{MJD(BTDB)} = 55618.926447(9) +\, 0.116\,015\,49(5) E,\]
which is consistent with the ephemeris of \citet{drake10}.

\begin{figure*}
\centering
\epsscale{.99}
\plotone{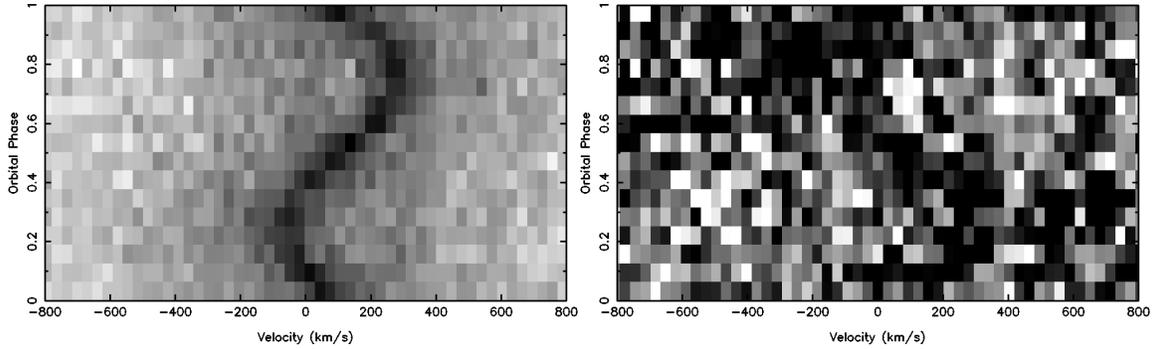}
\caption{\emph{Left:} Trailed spectrum of the H$\alpha$ line, the scale runs from black (50\% of the continuum level) to white (90\% of the continuum level). The absorption from the primary star is clearly visible. More subtly, absorption from the secondary star is visible moving in anti-phase with the primary stars absorption. \emph{Right:} Same as left hand panel but with the primary stars contribution removed (black is -2\% of the continuum, white is 4\% of the continuum). The absorption from the secondary star is now visible.\label{trail}}
\end{figure*}

\subsection{Temperatures}  \label{temps}

We used data from the Sloan Digital Sky Survey (SDSS) to determine the temperatures of the two white dwarfs in CSS 41177. Initially, we fitted its SDSS spectrum as a single white dwarf using the models of \citet{koester05}, which results in a $T_\mathrm{eff,1}=21,535\pm214$K and $\log g=7.36\pm0.44$. To determine the temperature of both stars better, we fitted the SDSS $ugriz$ magnitudes with a composite of two DA white dwarf models, with the additional constraint of the surface brightness ratio $S_1/S_2=2.90\pm0.23$ in the LT/RISE band, as measured from the light curve. This fit results in $T_\mathrm{eff,1}=21,100\pm800$K and $T_\mathrm{eff,2}=10,500\pm500$K, where the quoted errors are estimated, as they correlate with the uncertainties in the surface gravities. The best-fit composite model reproduces well the observed GALEX ultraviolet fluxes. 

\subsection{Radial velocities}

The left hand panel of Figure~\ref{trail} shows the trailed spectrum of CSS 41177 around the H$\alpha$ line. The non-LTE absorption core from the primary (hotter) star is clearly visible. There is also a weaker absorption component moving in anti-phase with the absorption from the primary star originating from the secondary (cooler) star. No other features are visible in the GMOS spectra.

Initially we fitted just the primary star's absorption. For each spectrum we fitted the line using a combination of a straight line plus a Gaussian. However, the radial velocity measured does not represent the true radial velocity of the primary star as the absorption from the secondary star causes a slight underestimation. Nevertheless, we used these velocities to correct out the motion of the primary star. We then averaged the shifted spectra to obtain the rest-frame spectrum of the primary star which we then subtracted from all the observed spectra. The result is shown in the right hand panel of Figure~\ref{trail}, clearly showing the absorption from the secondary star.

We fitted the absorption from the secondary star in the same way as the primary star. This too underestimated the true velocity since the primary star's contribution had not been completely removed. However, both fits provide us with an initial starting point for $K_1$ and $K_2$, the radial velocity amplitudes of both stars. In order to measure accurate radial velocities both absorption components need to be fitted simultaneously. We fit both lines together by simultaneously fitting all of the spectra. We used a combination of a straight line and Gaussians for each spectrum (including a broad Gaussian component to account for the wings of the primary star's absorption) and allowed the position of the Gaussians to change velocity according to
\begin{eqnarray}
V = \gamma + K\sin(2 \pi \phi), \nonumber
\end{eqnarray}
for each star, where $\gamma$ is the offset in the line from its rest wavelength and $\phi$ is the orbital phase of the spectrum. Due to the faintness of the secondary star's absorption we keep the offset $\gamma_2=\gamma_1$ to reduce the number of degrees of freedom. All other parameters were initially set to those measured from fitting the lines individually. We fitted the data using Levenberg-Marquardt minimisation \citep{press86}. The resultant fit had a reduced $\chi^2$ of 1.02. We find $K_1 = 177\pm3$ km$\,$s$^{-1}$, $K_2 = 181\pm20$ km$\,$s$^{-1}$ and $\gamma = 109\pm2$ km$\,$s$^{-1}$. These values imply that CSS 41177 is an equal mass binary (or very close to, given the uncertainty on K$_2$). 

\subsection{System Parameters}

\begin{figure*}
\centering
\epsscale{.99}
\plotone{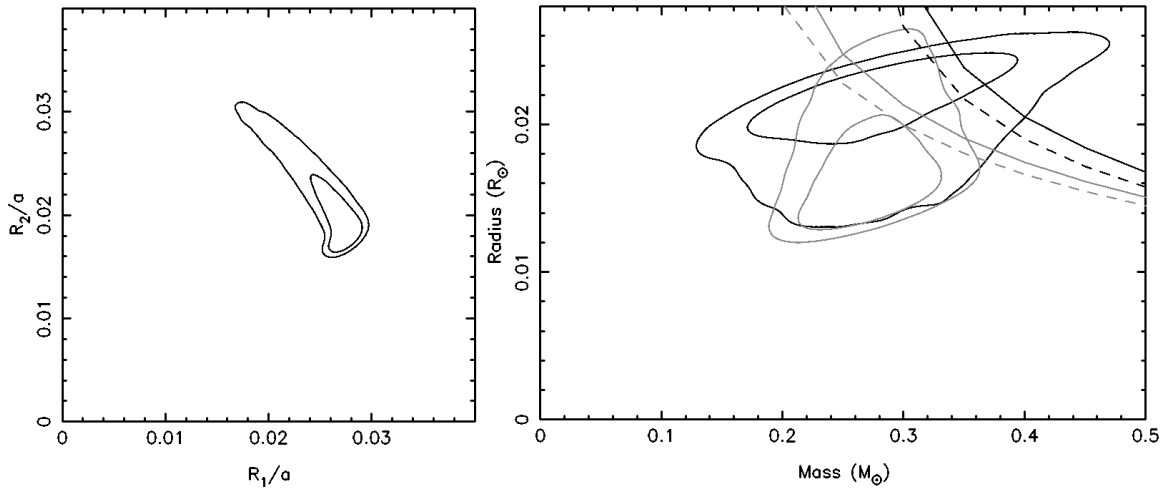}
\caption{\emph{Left:} The contours show the 68 and 95 percentile regions of the scaled radii $R_\mathrm{1}/a$ and $R_\mathrm{2}/a$, where $a$ is the orbital separation, constrained by the RISE photometry. \emph{Right:} The contours show the 68 and 95 percentile regions of the masses and radii for the hot (black lines) and cool (gray lines) white dwarfs. Also shown are mass-radius relations for helium core white dwarfs with hydrogen layer fractional masses of M$_\mathrm{H}/\mathrm{M}_*=10^{-4}$ (solid lines) and M$_\mathrm{H}/\mathrm{M}_*=10^{-8}$ (dashed lines) from \citet{benvenuto98} for temperatures of 10,500K and 21,100K. The confidence regions shown here are for two parameters jointly and should not be directly compared to the uncertainties in Table~\ref{params}, which are more closely related to single parameter confidence intervals.\label{constrain}}
\end{figure*}

The RISE photometry constrains $i$, $R_\mathrm{1}/a$ and $R_\mathrm{2}/a$. $K_1$ and $K_2$ are constrained from the spectroscopy, therefore we can determine the masses and radii of both white dwarfs using Kepler's third law,
\begin{eqnarray}
\frac{G(M_1 + M_2)}{a^3} = \frac{4 \pi^2}{P^2},
\end{eqnarray}
where $P$ is the orbital period, given the orbital separation from 
\begin{eqnarray}
\frac{2 \pi}{P}\, a\sin{i} = K_1 + K_2.
\end{eqnarray}
We determined the masses and radii for each model produced in the MCMC analysis of the light curve using the measured radial velocities and Kepler's laws. The right hand panel of Figure~\ref{constrain} shows the regions of allowed masses and radii for both white dwarfs based only upon the photometric and spectroscopic constraints. Also shown are models of helium core white dwarfs with hydrogen layer fractional masses of M$_\mathrm{H}/\mathrm{M}_*=10^{-4}$ (solid lines) and M$_\mathrm{H}/\mathrm{M}_*=10^{-8}$ (dashed lines) from \citet{benvenuto98} for temperatures of 10,500K (gray) and 21,100K (black) which, given the large uncertainties in our measurements, are consistent with the measured masses and radii. The final system parameters are listed in Table~\ref{params}.

\section{Conclusions}

\begin{table}
\begin{center}
\caption{Stellar and binary parameters for CSS 41177.\label{params}}
\begin{tabular}{@{}lccc@{}}
\tableline\tableline
Parameter & Value & Parameter & Value \\
\tableline
RA        & 10:05:59.11      & $K_1$ (km$\,$s$^{-1}$) & $177\pm3$  \\
Dec       & +22:49:32.3      & $K_2$ (km$\,$s$^{-1}$) & $181\pm20$ \\
GALEX fuv & $16.72\pm0.04$   & T$_\mathrm{eff,1}$ (K) & $21,100\pm800$ \\
GALEX nuv & $16.98\pm0.03$   & T$_\mathrm{eff,2}$ (K) & $10,500\pm500$ \\
$u$       & $17.317\pm0.010$ & M$_1$ (M$_{\sun}$) & $0.283\pm0.064$  \\
$g$       & $17.266\pm0.005$ & M$_2$ (M$_{\sun}$) & $0.274\pm0.034$  \\
$r$       & $17.612\pm0.006$ & R$_1$ (R$_{\sun}$) & $0.0210\pm0.0026$\\
$i$       & $17.899\pm0.008$ & R$_2$ (R$_{\sun}$) & $0.0174\pm0.0031$\\
$z$       & $18.151\pm0.024$ & a (R$_{\sun}$)     & $0.821\pm0.048$  \\
d (pc)    & $350\pm13$       & $i$ (deg)         & $89.2\pm0.3$   \\
P$_\mathrm{orb}$ (days)& 0.1160156(1) & & \\
\tableline
\end{tabular}
\end{center}
\end{table}

We have discovered that CSS 41177 is a detached, eclipsing, double white dwarf binary containing two low-mass helium core white dwarfs with masses of M$_1 = 0.283\pm0.064$M$_{\sun}$ and M$_2 = 0.274\pm0.034$M$_{\sun}$ and radii of R$_1 = 0.0210\pm0.0026$R$_{\sun}$ and R$_2 = 0.0174\pm0.0031$R$_{\sun}$. The hotter white dwarf has a temperature of $21,100\pm800$K and the cooler white dwarf has a temperature of $10,500\pm500$K placing it on the red edge of the instability strip. 

CSS 41177 is ideally set up to measure precise model-independent masses and radii for both white dwarfs and thus can be used to test white dwarf mass-radius relations. To date only one white dwarf has had its mass and radius independently measured to high precision (NN Ser, \citealt{parsons10}), therefore CSS 41177 offers two additional white dwarfs which can test mass-radius relations at the lower mass range.

Since the white dwarfs in CSS 41177 have roughly equal masses they will eventually undergo a merger resulting in the formation of a single $\sim0.6$M$_{\sun}$ sdB star \citep{han03}. This will occur in roughly 1.1 Gyr due to the loss of orbital angular momentum via gravitational radiation. The regular eclipse times from CSS 41177 should make it an excellent timing standard for optical astronomy, provided there are no unseen companions.

\acknowledgments

Balmer/Lyman lines in the models were calculated with the modified Stark broadening profiles of \citet{tremblay09}, kindly made available by the authors. TRM and BTG acknowledge support from the Science and Technology Facilities Council (STFC) grant number ST/F002599/1. The Liverpool Telescope is operated on the island of La Palma by Liverpool John Moores University in the Spanish Observatorio del Roque de los Muchachos of the Instituto de Astrofisica de Canarias with financial support from the UK Science and Technology Facilities Council. Based on observations obtained at the Gemini Observatory under program ID GN-2011A-DD-1, which is operated by the Association of Universities for Research in Astronomy, Inc., under a cooperative agreement with the NSF on behalf of the Gemini partnership: the National Science Foundation (United States), the Science and Technology Facilities Council (United Kingdom), the National Research Council (Canada), CONICYT (Chile), the Australian Research Council (Australia), Minist\'{e}rio da Ci\^{e}ncia e Tecnologia (Brazil) and Ministerio de Ciencia, Tecnolog\'{i}a e Innovaci\'{o}n Productiva (Argentina). 

{\it Facilities:} \facility{LT (RISE)}, \facility{Gemini (GMOS)}

\end{document}